\documentclass{llncs}

\newif\ifanonymousversion

\usepackage[utf8]{inputenc}
\usepackage{microtype}
\usepackage{hyperref}
\usepackage{cleveref}
\usepackage{tikz}
\usetikzlibrary{decorations.pathreplacing,calligraphy}
\usepackage{xcolor}
\usepackage{xspace}
\usepackage{subcaption}
\usepackage{utfsym}
\usepackage{booktabs}
\usepackage{float}
\usepackage{todonotes}
\usepackage{booktabs}
\usepackage{multirow}
\usepackage[title]{appendix}

\newcommand{\msg}[1]{\ensuremath{\mathtt{#1}}\xspace}

\newcommand{\id}{id} 

\newcommand{\CtoS}{C$\to$S\xspace}
\newcommand{\StoC}{S$\to$C\xspace}

\newcommand{\EncryptedHS}[1]{$\{$#1$\}$}
\newcommand{\EncryptedApp}[1]{$[$#1$]$}

\newcommand{\CHELO}{\msg{Client}\allowbreak{}\msg{Hello}}
\newcommand{\SHELO}{\msg{Server}\allowbreak{}\msg{Hello}}

\newcommand{\sCHELO}{\msg{CH}}
\newcommand{\sSHELO}{\msg{SH}}
\newcommand{\sENCEX}{\msg{ENCEXT}}
\newcommand{\sCERT}{\msg{CRT}}
\newcommand{\sCERTV}{\msg{CRTVER}}
\newcommand{\sCERTR}{\msg{CRTREQ}}
\newcommand{\sFIN}{\msg{FIN}}

\colorlet{udp}{green!40!black}
\colorlet{tcp}{blue!80!black}
\colorlet{tcptls}{magenta!80!black}
\colorlet{udptls}{blue!60!green}

\newcommand{\BobAction}[1]{
	\node[] at (\BobX, \CurrentY) {#1};
}

\newcommand{\AliceToBob}[3][->]{
	\NextLine[0.5]
	\draw[#1] (\AliceBobArrow,\CurrentY) -- node[above,yshift=-0.7mm] {#2} node[below] {#3} (\BobAliceArrow,\CurrentY) ;
}
\newcommand{\BobToAlice}[3][->]{
	\NextLine[0.5]
	\draw[#1] (\BobAliceArrow,\CurrentY) -- node[above,yshift=-0.7mm] {#2} node[below] {#3} (\AliceBobArrow,\CurrentY) ;
}
\newcommand{\AliceToCharlie}[3][->]{
	\NextLine[0.5]
	\draw[#1] (\AliceCharlieArrow,\CurrentY) -- node[above,yshift=-0.7mm] {#2} node[below] {#3} (\CharlieAliceArrow,\CurrentY) ;
}
\newcommand{\CharlieToAlice}[3][->]{
	\NextLine[0.5]
	\draw[#1] (\CharlieAliceArrow,\CurrentY) -- node[above,yshift=-0.7mm] {#2} node[below] {#3} (\AliceCharlieArrow,\CurrentY) ;
}

\newcommand{\NextLine}[1][1.0]{
	\pgfmathparse{\CurrentY+#1}
	\edef\CurrentY{\pgfmathresult}
}

\edef\BraceStartY{0}
\edef\BraceEndY{0}
\newcommand{\BraceStart}{\edef\BraceStartY{\CurrentY}}
\newcommand{\BraceEnd}{\edef\BraceEndY{\CurrentY}}
\newcommand{\Brace}[2]{
	\draw [decorate,decoration={brace},thick] (#1,\BraceStartY-0.25)--(#1,\BraceEndY+0.1) node[right=5pt,pos=0.5] {#2};
}

\newcommand{\pie}[1]{%
\begin{tikzpicture}
 \draw (0,0) circle (0.8ex);\fill (0.8ex,0) arc (0:#1:0.8ex) -- (0,0) -- cycle;
\end{tikzpicture}%
}
\newcommand{\circleFull}{\pie{360}\xspace}
\newcommand{\circleThrQ}{\pie{270}\xspace}
\newcommand{\circleHalf}{\pie{180}\xspace}
\newcommand{\circleQuar}{\pie{90}\xspace}
\newcommand{\circleZero}{\pie{0}\xspace}

\newcommand{\NA}{---}
 
\newcommand{\myparagraph}[1]{\medskip\noindent\textit{#1.} }

\title{\textbf{TurboTLS: TLS connection establishment with\\1 less round trip}}
\ifanonymousversion
\author{Anonymous submission to CT-RSA}
\else
\author{Carlos Aguilar-Melchor$^{1}$, Thomas Bailleux$^{1}$, Jason Goertzen$^{1,2}$, Adrien Guinet$^{1}$,\\ David Joseph$^{1}$, and Douglas Stebila$^{2}$}
\institute{
SandboxAQ\\\emph{Palo Alto, USA}\\
\email{\{firstname.lastname\}@sandboxaq.com}\and
University of Waterloo\\\emph{Waterloo, Canada}\\
\email{\{firstname.lastname\}@uwaterloo.ca}
}
\fi

\begin{document}

\maketitle

\begin{abstract}
We show how to establish TLS connections using one less round trip.  In our approach, which we call TurboTLS, the initial client-to-server and server-to-client flows of the TLS handshake are sent over UDP rather than TCP.  At the same time, in the same flights, the three-way TCP handshake is carried out.  Once the TCP connection is established, the client and server can complete the final flight of the TLS handshake over the TCP connection and continue using it for application data.  No changes are made to the contents of the TLS handshake protocol, only its delivery mechanism.  We avoid problems with UDP fragmentation by using \emph{request-based fragmentation}, in which the client sends in advance enough UDP requests to provide sufficient room for the server to fit its response with one response packet per request packet.  Clients can detect which servers support this without an additional round trip, if the server advertises its support in a DNS HTTPS resource record.  Experiments using our software implementation show substantial latency improvements.  On reliable connections, we effectively eliminate a round trip without any noticeable cost.  To ensure adequate performance on unreliable connections, we use lightweight packet ordering and buffering; we can have a client wait a very small time to receive a potentially lost packet (e.g.  a fraction of the RTT observed for the first fragment) before falling back to TCP without any further delay, since the TCP connection was already in the process of being established.  This approach offers substantial performance improvements with low complexity, even in heterogeneous network environments with poorly configured middleboxes.
 \end{abstract}

\section{Introduction}\label{sec:intro}

The Transport Layer Security (TLS) protocol is ubiquitous and provides security services to many network applications.  TLS runs over TCP.  As shown in \Cref{fig:tls}, the main flow for TLS 1.3 connection establishment \cite{rfc8446} in a web browser is as follows.

\begin{figure*}[t]
\begin{minipage}{0.48\textwidth}
	\centering
        \scalebox{0.7}{
\begin{tikzpicture}[yscale=-1,>=latex,
	every node/.style={
		font=\scriptsize,
		align=center
	},
	udpflow/.style={udp,dashed,->,thick},
	tcpflow/.style={tcp,->,thick},
	tcptlsflow/.style={tcptls,->,thick}
]

\edef\AliceX{0}
\edef\BobX{5.5}
\edef\CharlieX{7}
\edef\AliceBobArrow{0}
\edef\BobAliceArrow{5.5}
\edef\AliceCharlieArrow{0}
\edef\CharlieAliceArrow{7}
\edef\CurrentY{0}

\node [rectangle,draw,inner sep=5pt,right] at (\AliceX,\CurrentY) {\textbf{TLS Client}};
\node [rectangle,draw,inner sep=5pt,left] at (\CharlieX,\CurrentY) {\textbf{DNS Server}};

\NextLine[0.25]
\AliceToCharlie[udpflow]{DNS: A request}{}
\AliceToCharlie[udpflow]{DNS: AAAA request}{}
\AliceToCharlie[udpflow]{DNS: HTTPS RR request}{}
\CharlieToAlice[udpflow]{DNS: A response}{}
\CharlieToAlice[udpflow]{DNS: AAAA response}{}
\CharlieToAlice[udpflow]{DNS: HTTPS RR response}{}

\NextLine[0.7]
\node [rectangle,draw,inner sep=5pt,right] at (\AliceX,\CurrentY) {\textbf{TLS Client}};
\node [rectangle,draw,inner sep=5pt,left] at (\BobX,\CurrentY) {\textbf{TLS Server}};

\NextLine[0.25]
\AliceToBob[tcpflow]{TCP: SYN}{} \BraceStart
\BobToAlice[tcpflow]{TCP: SYN ACK}{} \BraceEnd
\Brace{\BobX+0.5}{1st \\ RTT}
\NextLine[0.5]
\AliceToBob[tcpflow]{TCP: ACK}{} \BraceStart
\AliceToBob[tcptlsflow]{TCP: TLS \sCHELO}{}
\BobToAlice[tcptlsflow]{TCP: TLS \sSHELO,\EncryptedHS{\sENCEX,$\sCERTR^*$,$\sCERT^*$,$\sCERTV^*$,$\sFIN$}}{}
\BobToAlice[tcptlsflow]{TCP: TLS \EncryptedApp{Application Data$^*$}}{} \BraceEnd
\Brace{\BobX+0.5}{2nd \\ RTT}
\NextLine[0.5]
\AliceToBob[tcptlsflow]{TCP: TLS \EncryptedHS{$\sCERT^*$,$\sCERTV^*$,$\sFIN$}}{} \BraceStart
\AliceToBob[tcptlsflow]{TCP: TLS \EncryptedApp{Application Data$^*$}}{} \BraceEnd
\Brace{\BobX+0.5}{3rd \\ RTT}
\end{tikzpicture}
}
	\caption{TLS 1.3 connection establishment}
	\label{fig:tls}
\end{minipage}
\begin{minipage}{0.48\textwidth}
	\centering
        \scalebox{0.7}{
\begin{tikzpicture}[yscale=-1,>=latex,
	every node/.style={
		font=\scriptsize,
		align=center
	},
	udpflow/.style={udp,dashed,->,thick},
	tcpflow/.style={tcp,->,thick},
	tcptlsflow/.style={tcptls,->,thick},
	udptlsflow/.style={udptls,dashed,->,thick}
]

\edef\AliceX{0}
\edef\BobX{5.5}
\edef\CharlieX{7.5}
\edef\AliceBobArrow{0}
\edef\BobAliceArrow{5.5}
\edef\AliceCharlieArrow{0}
\edef\CharlieAliceArrow{7.5}
\edef\CurrentY{0}

\node [rectangle,draw,inner sep=5pt,right] at (\AliceX,\CurrentY) {\textbf{TLS Client}};
\node [rectangle,draw,inner sep=5pt,left] at (\CharlieX,\CurrentY) {\textbf{DNS Server}};

\NextLine[0.25]
\AliceToCharlie[udpflow]{DNS: A request}{}
\AliceToCharlie[udpflow]{DNS: AAAA request}{}
\AliceToCharlie[udpflow]{DNS: HTTPS RR request}{}
\CharlieToAlice[udpflow]{DNS: A response}{}
\CharlieToAlice[udpflow]{DNS: AAAA response}{}
\CharlieToAlice[udpflow]{DNS: HTTPS RR response w/TurboTLS flag}{}

\NextLine[0.7]
\node [rectangle,draw,inner sep=5pt,right] at (\AliceX,\CurrentY) {\textbf{TLS Client}};
\node [rectangle,draw,inner sep=5pt,left] at (\BobX,\CurrentY) {\textbf{TLS Server}};

\NextLine[0.25]
\AliceToBob[udptlsflow]{UDP: TurboTLS $\id$, TLS \sCHELO frag \#1}{} \BraceStart
\AliceToBob[udptlsflow]{UDP: TurboTLS $\id$, TLS \sCHELO frag \#2}{}
\AliceToBob[udptlsflow]{UDP: TurboTLS $\id$, empty frag \#1}{}
\AliceToBob[udptlsflow]{UDP: TurboTLS $\id$, empty frag \#2}{}
\AliceToBob[tcpflow]{TCP: SYN}{}
\NextLine[0.7]
\BobAction{Compute TLS \\ response \sSHELO, \\ \EncryptedHS{\sENCEX,$\sCERTR^*$, \\ $\sCERT^*$,$\sCERTV^*$,$\sFIN$}}
\NextLine[0.6]
\BobToAlice[udptlsflow]{UDP: TurboTLS $\id$, TLS response frag \#1}{}
\BobToAlice[udptlsflow]{UDP: TurboTLS $\id$, TLS response frag \#2}{}
\BobToAlice[udptlsflow]{UDP: TurboTLS $\id$, TLS response frag \#3}{}
\BobToAlice[tcpflow]{TCP: SYN ACK}{} \BraceEnd
\Brace{\BobX+1.25}{1st \\ RTT}

\NextLine[0.5]
\AliceToBob[tcpflow]{TCP: ACK}{} \BraceStart
\AliceToBob[tcptlsflow]{TCP: TurboTLS $\id$, TLS \EncryptedHS{$\sCERT^*$,$\sCERTV^*$,$\sFIN$}}{}
\AliceToBob[tcptlsflow]{TCP: TLS \EncryptedApp{Application Data$^*$}}{} \BraceEnd
\Brace{\BobX+1.25}{2nd \\ RTT}

\end{tikzpicture}
}
	\caption{TurboTLS connection establishment}
	\label{fig:turbotls}
\end{minipage}

\caption[Comparing message flow of TLS 1.3 and TurboTLS]{Comparing message flow of TLS 1.3 and TurboTLS. Legend: dashed line - UDP; solid line - TCP.  $^*$~denotes optional message.  $\{\dots\}$ denotes messages encrypted using TLS handshake traffic secret and $[\dots]$ denotes messages encrypted using TLS application traffic secret.}
\label{fig:tls_vs_turbotls}
\end{figure*}

First, the client makes a DNS query to translate the provided domain name into an IP address.  Modern browsers simultaneously request from the DNS server an HTTPS resource record \cite{ietf-dnsop-svcb-https-11} which can provide additional information about the server's HTTPS configuration.  Next, the client performs the TCP three-way handshake with the server.  Once the TCP handshake has completed and a TCP connection is established, the TLS handshake can begin; it requires one client-to-server (\CtoS) flow and one server-to-client (\StoC) flow before the client can start sending application data.  

In total, excluding the DNS resolution, this results in two round trips before the client can send its first byte of application data (the TCP handshake and the first \CtoS and \StoC flows of the TLS handshake), and another round trip before the client receives its first byte of response.

TLS does have a pre-shared key mode that allows for an abbreviated handshake permitting application data to be sent in the first \CtoS TLS flow, but this requires that the client and server have a pre-shared key in advance, established either through some out-of-band mechanism or saved from a previous TLS connection for session resumption. 

\myparagraph{Our contributions}
We describe a method, which we call TurboTLS, for removing one round trip of latency from TLS connection establishment by transmitting the first two flows of the TLS handshake over UDP while doing the TCP three-way handshake in parallel, then switching over to the TCP connection for the final \CtoS handshake flow and the transmission of application data.  
The message flow of TurboTLS is shown in \Cref{fig:turbotls}.  
It allows the client to start sending its first byte of application in just one round trip (excluding the DNS resolution), without requiring any pre-shared key.
TurboTLS does not require any change to the contents or state machine of the TLS protocol: it only changes the network delivery mechanism.
We employ several techniques to make TurboTLS operate smoothly in a heterogeneous network environment where there may be UDP packet loss, where some servers may not support TurboTLS, and where intermediary network devices may have trouble with UDP fragmentation. 
We implement TurboTLS and compare its network performance with TLS 1.3 and QUIC over local, national, and intercontinental distance connections, as shown in \Cref{fig:results} and \Cref{tab:results}.  In general, TurboTLS achieves median latency approximately the same as QUIC, which is a 50\% improvement compared to TLS 1.3 in long-distance connections, and slightly less in short-distance connections to fixed overhead.

\Cref{tab:comparison} summarizes the characteristics of TurboTLS compared with other relevant network security protocols; see \Cref{sec:background} and \Cref{sec:discussion:comparison} for more details.

\begin{table*}
\begin{center}
\resizebox{\textwidth}{!}{
\begin{tabular}{lcccccccc} 
\toprule
& \textbf{Runs}
& \textbf{UDP 1 req.} 
& \textbf{Provides} 
& \textbf{Kernel}
& \textbf{No}
& \textbf{TLS-}
& \textbf{Widely}
& \textbf{RTT to}
\\
& \textbf{over}
& \textbf{$\Rightarrow$ 1 resp.}
& \textbf{conn.}
& \textbf{netw.}
& \textbf{state}
& \textbf{based}
& \textbf{deployed}
& \textbf{1st byte}
\\
\midrule
TLS 1.2 & TCP & 
\NA         & \circleFull & \circleFull & \circleFull & \circleFull & \circleFull & 3 \\
TLS 1.2 FalseStart & TCP & 
\NA         & \circleFull & \circleFull & \circleFull & \circleFull & \circleFull & 2 \\
TLS 1.3 & TCP & 
\NA         & \circleFull & \circleFull & \circleFull & \circleFull & \circleFull & 2 \\
TLS 1.3 PSK & TCP & 
\NA         & \circleFull & \circleFull & \circleZero & \circleFull & \circleFull & 1 \\
TLS 1.3 ECH & TCP & 
\NA         & \circleFull & \circleFull & \circleZero & \circleFull & \circleHalf & 1 \\
OPTLS & TCP & 
\NA         & \circleFull & \circleFull & \circleZero & \circleFull & \circleZero & 1 \\
TLS 1.3 + TCP Fast Open & TCP &
\NA         & \circleFull & \circleFull & \circleZero & \circleFull & \circleQuar & 1 \\
DTLS 1.3 & UDP & 
\circleFull & \circleZero & \circleFull & \circleFull & \circleFull & \circleFull & 2 \\
QUIC & UDP & 
\circleZero & \circleFull & \circleZero & \circleFull & \circleThrQ & \circleHalf & 1 \\
MinimaLT & UDP & 
\circleFull & \circleFull & \circleZero & \circleFull & \circleZero & \circleZero & 1 \\
MinimaLT with state & UDP & 
\circleFull & \circleFull & \circleZero & \circleZero & \circleZero & \circleZero & 0 \\
\midrule
TurboTLS & UDP+TCP & 
\circleFull & \circleFull & \circleThrQ & \circleFull & \circleFull & \NA         & 1 \\
TurboTLS + PSK & UDP+TCP & 
\circleFull & \circleFull & \circleThrQ & \circleZero & \circleFull & \NA         & 0 \\
TurboTLS + ECH & UDP+TCP & 
\circleFull & \circleFull & \circleThrQ & \circleZero & \circleFull & \NA         & 0 \\
\bottomrule
\end{tabular}
}
\end{center}
\caption{Characteristics of TurboTLS compared to TLS and other optimized/accelerated protocols and variants.}
\label{tab:comparison}
\textbf{Legend:} \circleFull: yes; \circleThrQ, \circleHalf, \circleQuar: partial; \circleZero: no; \NA: not applicable.
\textbf{Columns:} UDP 1 req. $\Rightarrow$ 1 resp.: does each UDP request packet lead to at most one response packet?
Provides conn.: does the protocol provide connection-oriented (reliable, in-order) transport to the application?
Kernel netw.: are connection-oriented features generally implemented in the kernel?
No state: can full optimization be achieved without pre-shared state between client and server?
RTT to 1st byte: how many round trips required until the client can send its first application byte, including TCP 3-way handshake if necessary.
\end{table*}

While our primary focus is on TLS connection establishment over TCP, many other network protocols have a similar structure.
Our techniques can be generalized to an approach we call ``turbo transport'' which accelerates connection establishment by shifting portions of the connection establishment to UDP.
\section{Background}\label{sec:background}

An application wishing to establish a secure connection between a client and a server will select a protocol, or combination of protocols across network layers, depending on a number of factors. 
Depending on the application (performance requirements, reliability requirements, etc.) there may be a preference for a connection-oriented or connectionless protocol.
Finally, if a server does not support a protocol, or an initial request is blocked due to other reasons such as firewall filtering, then the client may need to fall back to another protocol that reaches the server and is supported.

In this section, we review several existing options to set up application-level secure channels, focusing on the TLS protocol, variants of TLS, and protocols aiming to replace TLS.
Our first-level categorization is whether the protocol runs over TCP or UDP.

\subsection{Secure channel protocols over TCP}\label{sec:background:tcp}

Applications requiring connection-oriented communication typically run over TCP, such as `vanilla' TLS. 
TCP uses an initial round trip to set up the connection,  using the TCP three-way handshake, then a further round trip is needed to complete the TLS 1.3 handshake (or two further round trips for TLS 1.2 \cite{rfc5246}), during which the cryptographic parameters are negotiated, session keys are exchanged, and authentication happens.

While the number of rounds trips and the resulting inherent latency is not always a problem for clients/servers in close proximity to one another, this presents a significant inconvenience where parties are far apart or suffer high network latency. To ameliorate this issue, a series of optimizations to TLS have been proposed, using a range of approaches. 

\myparagraph{Data-based optimizations}
Some optimizations reduce the amount of data transmitted without reducing the number of round trips. Perhaps the simplest approach is that of Compact TLS \cite{ietf-tls-ctls-06}, which changes the format of TLS handshake messages by removing obsolete fields and defining profiles of common options. Another light-touch optimization is TLS Cached Information Extension \cite{rfc7924}, which allows clients and servers to indicate they already have certain sets of values, such as intermediate certificates, to avoid re-transmitting them, which can save a significant amount of bandwidth, especially in the context of post-quantum cryptography which typically results in larger intermediate certificates \cite{NDSS:SikKamDev20}.

\myparagraph{Optimizations using previous state}
Some optimizations are possible if the client has some prior server-dependent state, either from a previous connection or from some public directory.

TLS has a pre-shared key (PSK) mode in which a client can make use of a pre-shared symmetric key to save one round trip, allowing a client to start sending application data in the first TLS 1.3 PSK flow (and thus on the second \CtoS flow including the TCP three-way handshake).

TCP Fast Open \cite{rfc7413} allows a client to save a cryptographic cookie from a previous TCP connection and use it in a subsequent TCP connection to immediately start sending application data without having to do a TCP three-way handshake on the subsequent connection. TLS running over TCP Fast Open would obviously then save one round trip.

OPTLS \cite{EUROSP:KraWee16} was an alternative design for the TLS 1.3 handshake, running over TCP, which supported a so-called 0-RTT mode allowing for a client to send application data in its first \CtoS TLS flow (and thus on the second \CtoS flow including the TCP three-way handshake) provided that the client had previously cached or obtained out-of-band the server's semi-static public key.  

Encrypted Client Hello \cite{ietf-tls-esni-15} is a proposed TLS extension that enables the client to encrypt more of the \CHELO message as well as send early application data in the first TLS \CtoS flight (and thus on the second \CtoS flow including the TCP three-way handshake) provided the client (similarly to in OPTLS) has previously cached or obtained out-of-band a public key of the server (which could be distributed in a DNS record).

There were also several other modifications to TLS 1.2 that made use of previous state, including TLS Snap Start \cite{agl-tls-snapstart-00} and ``fast-track'' client-side caching \cite{TISSEC:ShaBonRes04}.

\subsection{Secure channel protocols over UDP}\label{sec:background:udp}

Another branch of optimizations utilizes the connectionless properties of UDP to fast-track performance. 
However, since many applications need to connection-oriented channels for data transmissions, most optimizations running on top of UDP specify their own procedures for packet reordering, packet loss, and session management, although we first briefly discuss DTLS, which does not.

\myparagraph{Protocols not providing connection-oriented features to applications}
DTLS \cite{rfc9147} runs exclusively over UDP, including for transmission of application data. Because it runs on UDP, it is left to the application to reorder packets and deal with loss. Cryptographically, DTLS is based on TLS.  DTLS can be particularly useful when trying to avoid problems such as TCP meltdown, whereby applications may be trying to transport TCP traffic inside a secure tunnel which also runs on TCP, essentially stacking TCP upon TCP and thereby amplifying the occurrence of TCP timeout and other related problems. For this reason DTLS is often used for VPN applications. 

\myparagraph{Protocols providing connection-oriented features to applications}
MinimaLT \cite{CCS:PZSBL13} runs exclusively over UDP and uses a completely different protocol design compared to TLS, but has not, to date, seen widespread usage.  

QUIC \cite{rfc9000} is another approach. Designed originally to improve performance of encrypted transport for Google's internal services, QUIC is an ambitious and completely separate connection-oriented protocol running on top of UDP. 
Like MinimaLT, QUIC fundamentally merges the transport and security layers, and provides many other protocol-specific optimizations, such as providing varying header lengths (a longer header format is used for packets establishing connections), ACK-based packet loss detection which overcomes the instability of UDP by providing a grace period to in-flight packets, packet re-ordering, and others. 
A further benefit of QUIC is that for re-established connections, it is possible to send encrypted application data in the first packet by re-using previously agreed cryptographic parameters and utilizing a pre-shared key setup, using a technique similar to those of OPTLS, and again at the cost of forward secrecy for the initial data sent.

\section{TurboTLS design}\label{sec:design}

As described in \Cref{fig:turbotls}, TurboTLS sends part of the TLS handshake over UDP, rather than TCP.
Switching from TCP to UDP for handshake establishment means we cannot rely on TCP's features, namely connection-oriented, reliable, in-order delivery.  
However, since the rest of the connection will still run over TCP and only part of the handshake runs over UDP,
we can reproduce the required functionality in a lightweight way without adding latency and allowing for a simple implementation.

\myparagraph{Fragmentation}
One of the major problems to deal with is that of fragmentation.  TLS handshake messages can be too large to fit in a single packet -- especially with long certificate chains or if post-quantum algorithms are used.  

Obviously the client can fragment its first \CtoS flow across multiple UDP packets.  To allow a server to link fragments received across multiple UDP requests, we add a 12-byte connection identifier field, containing a client-selected random value $\id$ that is used across all TurboTLS fragments sent by the client. The connection identifier is also included in the first message on the established TLS connection to allow the server to link together data received on the UDP and TCP connections. To allow the server to reassemble fragments if they arrive out-of-order, each fragment includes the total length of the original message as well as the offset of the current fragment; this can allow the server to easily copy fragments into the right position within a buffer as they are received.

Similarly, the server can fragment its first \StoC flow across multiple UDP packets.  One additional problem here however is that the \StoC flow is typically larger than the \CtoS flow (as it typically contains one or more certificates), so the server may have to send more UDP response packets than UDP request packets.  As noted by \cite{song-atr-large-resp-03} in the context of DNSSEC, many network devices do not behave well when receiving multiple UDP responses to a single UDP request, and may close the port after the first packet, dropping the request.  Subsequent packets received at a closed port lead to ICMP failure alerts, which can be a nuisance.

We employ a recent method proposed by Goertzen and Stebila \cite{arxiv.2211.14196} for DNSSEC: request-based fragmentation.  In the context of large resource records in DNSSEC, \cite{arxiv.2211.14196} had the first response be a truncated response that included information about the size of the response, and then the client sent multiple additional requests, in parallel, for the remaining fragments.  This ensured that there was only one UDP response for each UDP request.  We adapt that method for TurboTLS: the client, in its first \CtoS flow, fragments its own \CtoS data across multiple UDP packets, and additionally sends (in parallel) enough nearly-empty UDP requests for a predicted upper bound on the number of fragments the server will need to fit its response.  This preserves the model of each UDP request receiving a single UDP response, reducing the impact of misbehaving network devices and also reducing the potential for DDoS amplification attacks.

\myparagraph{Reliability}
UDP does not have reliable delivery, so packets may be lost.  Since the first TurboTLS round-trip includes the TCP handshake, we can immediately fall back to TCP if a UDP packet is lost in either direction.  This will induce a latency cost of however long the client decides to wait for UDP packets to arrive before giving up and assuming they were lost.

In an implementation, the client delay could be a fixed number of milliseconds, or could be variable depending on observed network conditions; this need not be fixed by a standard.
We believe that in many cases a client delay of just 2ms after the TCP reply is received in the first round trip will be enough to ensure UDP responses are received a large majority of the time.  In other words, by tolerating a potential 2ms of extra latency on $X$\% of connections, we can save an entire round-trip on a large proportion ($100-X$\%) of the connections. 
This mechanic was not implemented in the experimental results presented here and constitutes future work.

\myparagraph{Advertising support}
To protect servers who do not support TurboTLS from being bombarded with unwanted UDP traffic, it would be preferable if clients only used TurboTLS with servers that they already know support it.  Clients could cache this information from previous non-TurboTLS connections, but in fact we can do better.  Even on the first visit to a server, we can communicate server support for TurboTLS to the client, without an extra round trip, using the HTTPS resource record in DNS \cite{ietf-dnsop-svcb-https-11}.  Today when web browsers perform the DNS lookup for the domain name in question, they typically send three requests in parallel: an A query for an IPv4 address, an AAAA query for an IPv6 address, and a query for an HTTPS resource record \cite{ietf-dnsop-svcb-https-11}.  Servers can advertise support for TurboTLS with an additional flag in the HTTPS resource record and clients can check for it without incurring any extra latency.

\section{Features and advantages}\label{sec:features}

\subsection{Comparison with other protocols}\label{sec:discussion:comparison}

The protocols presented in \Cref{sec:background} fall in one or more of the following categories: 
\begin{itemize}
\item doing more than one round trip (TLS 1.2, TLS 1.2 FalseStart, TLS 1.3, DTLS 1.3, Compact TLS and TLS Cached Information Extension);
\item modifying TCP/UDP directly or modify the way TCP/UDP are expected to be used (TLS 1.3 + TCP Fast Open, DTLS, MinimaLT, QUIC); or
\item maintaining a state (TLS 1.3 + TCP Fast Open, TLS 1.3 PSK, OPTLS, TLS Encrypted Client Hello).
\end{itemize}

TurboTLS requires one round trip, uses TCP and UDP without modifying them, and as middleboxes would expect them to be used, and does not require a state. The rest of this section is dedicated to explaining the drawbacks of falling into one of the three categories above. We include a comparison with alternative protocols, but for a more detailed comparison with QUIC, which is gaining significant traction, we refer the reader to Section~\ref{sec:usecases}.

\smallskip 

\myparagraph{One round trip}
Doing more than one round trip increases latency by at least one RTT. As already noted, global studies provided to the community by CAIDA \cite{CAIDA} show that the RTT for median connections is between 50 and 200ms, mainly due to hops and distance, and if we consider the 90th percentile of connections, RTTs are beyond 500ms. This latency introduction is amplified by an integer factor for protocols in which connections occur sequentially (e.g. get a web page, get frames in the page, get images in the frames). Besides user experience, this also has an impact on usual implementations in which a server thread from a pool will not come back to the pool until it finishes dealing with a client. For connections with an RTT over a few milliseconds, when replacing TLS 1.3 by TurboTLS, we indeed observed a multiplication by two of the maximum handshakes per second that could be handled by a simple, yet usual, server implementation (thread pool with as many threads as cores that asynchronously handles connections).

\myparagraph{Standard and expected usage of TCP/UDP}
Some protocols modify TCP/UDP or the way they are expected to be used. Of course, TCP Fast Open modifies TCP itself by introducing cookies for the first flight (which also requires a state). This probably explains why, even if the initial proposal is from 2011, it is still an experimental RFC and not enabled by default on most browsers. Other protocols, like DTLS, MinimaLT or QUIC, just use UDP, without modifying it, but do long-term bidirectional exchanges, which is not the usual for UDP. Long-term bidirectional exchanges are in general done over TCP, and most protocols using UDP either follow the one query/one reply model (e.g. DNS) or the one query/many replies model (e.g. FTP download). 

Using UDP for long-term bidirectional exchanges introduces two issues: instability and computational overhead. The main and simplest reason for UDP instability are firewall rules which often block such traffic, except for the one query/one reply model. Besides that, some middlebox functions for long term bidirectional traffic are only available for TCP and with UDP will either reduce performance or cause instability. For example, in the context of QUIC, extensive guidance \cite{rfc9312} is required to deal with usual traffic management over QUIC, with the RFC addressing issues such as: ``Passive Network Performance Measurement and Troubleshooting; Stateful Treatment of QUIC Traffic; Address Rewriting to Ensure Routing Stability; Server Cooperation with Load Balancers; Filtering Behavior; UDP Blocking, Throttling, and NAT Binding; DDoS Detection and Mitigation; Quality of Service Handling and ECMP Routing;'' and more. UDP bidirectional long-term exchanges that run through middleboxes that have not implemented these features (adapted to QUIC or to other protocols) will suffer from a lower quality of service and stability, sometimes with catastrophic effects~\cite{MiddleboxesVsQuic}. On top of that, when connection-oriented features are provided (e.g. by QUIC or MinimaLT), one general drawback is that the implementation of the protocol needs to provide for packet reordering and recovery from packet loss, in user-space, whereas protocols running over TCP receive that for free from the operating system's kernel-space TCP implementation, which has typically been highly tuned over many years, leading among other things to fewer interrupts and copies.

\myparagraph{No state}
Maintaining a state brings obvious issues (no benefit on first connection, lifetime, complexity) but most importantly, in the case of TLS, it also induces in general a loss of forward secrecy and thus of security.  
TLS 1.3 in pre-shared key (PSK) mode gains one round-trip only when it completely drops forward secrecy (as it relies on a pre-shared secret). This is somewhat mitigated by OPTLS and the Encrypted Client Hello TLS extension: if the client has previously obtained the server public key, then use of OPTLS and Encrypted Client Hello TLS extension are improved by one round-trip, and the loss of forward secrecy only affects the first flow of messages from the client to the server. The rest of the communication has forward secrecy. 

\myparagraph{No TLS changes}
Additionally, TurboTLS shares, with TLS 1.3 + TCP Fast Open, another nice feature: TurboTLS makes no change whatsoever to the content of a TLS handshake, only changes the delivery mechanism.  As a result, all cryptographic properties of TLS are untouched.  In fact, it is possible to implement TurboTLS without changing the client or server's TLS library at all, and instead use transparent proxies on both the client and server side to change the network delivery from pure TCP in TLS to UDP+TCP in TurboTLS. Of course in such a construction the initial client or server, who does not know TurboTLS, will observe two round trip times, but if each proxy is close to its host (say on the same machine), then the two round trip times will be negligible, and the higher latency client--server distance will only be covered over one round trip.

\subsection{Denial-of-Service (DoS) considerations}\label{sec:discussion:DoS}

We now consider the implications for TurboTLS of various types of denial-of-service and distributed denial-of-service attacks, including whether a TurboTLS server is a victim in a DoS attack or being leveraged by an attacker to direct a DDoS attack elsewhere. TurboTLS runs on top of both TCP and UDP so we have to consider attacks involving both protocols. 

\myparagraph{DoS attacks on TurboTLS servers}
The most significant TCP DoS attack is the SYN flood attack where a target machine is overwhelmed by TCP SYN messages faster than it can process them. This is because a server, upon receiving a SYN, typically stores the source IP, TCP packet index number, and port in a `SYN queue', and this represents a half-open connection. An attacker could flood the server with SYN messages thereby exhausting its memory. The server cannot just arbitrarily drop connections because then legitimate users may find themselves unable to connect. There are many protections against SYN flood attacks, one of which is allocating only very small amounts (micro blocks) of memory to half-open connections. Another is using TCP cryptographic cookies \cite{syncookies,rfc6013} whereby the sequence number of the ACK encodes information about the SYN queue entry so that the server can reconstruct the entry even if it was not stored due to having a full SYN queue. TCP cookies enjoy support in the Linux kernel -- this and other such mitigations are already sufficient to protect TurboTLS from SYN floods.

In general there are several vectors to consider for resource exhaustion attacks on a server running TurboTLS.  
The server needs to maintain a buffer of received UDP packets containing fragments of a TLS \CHELO message.
To avoid memory exhaustion attacks, a server can safely bound the memory allocated to this buffer and flush old entries on a regular basis (e.g. after two seconds).
In the worst case, a legitimate client whose UDP packets are rejected from a busy server or flushed early will be able to fall back to vanilla TLS over TCP, and will incur negligible latency loss (compared to TLS over TCP) in doing so, because TurboTLS starts the TCP handshake in parallel to the first \CtoS UDP flow.
An attacker spoofing IP addresses and sending well-formed \CHELO messages could also try to exhaust a server's CPU resources by causing a large amount of cryptographic computation.
Again, a server under attack can limit the CPU resources allocated to UDP-received \CHELO messages, and then fall back to vanilla TLS over TCP.
In the worst case, legitimate clients affected by this and having to fall back to vanilla TLS over TCP will incur negligible latency loss compared to TLS over TCP since the TCP handshake has already been started in parallel.

\myparagraph{DDoS attacks leveraging TurboTLS servers}
UDP reflection attacks present another threat. Typical defenses against these are blocking unused ports, rate limiting based on expected traffic loads from peers (exorbitant traffic loads are likely to be malicious), or blocking IPs of other known vulnerable servers. However such defenses are provided by middleboxes and therefore do not affect the protocol. 

It should be noted here that the redundant UDP packets sent along with \CHELO are part of the TurboTLS-specific technique we call request-based-fragmentation to mitigate \textit{against} a client's middlebox defenses incorrectly filtering TurboTLS connections, as otherwise multiple UDP responses to a single UDP request could be flagged as malicious behaviour. Furthermore, the one-to-oneness of the UDP request/response significantly reduces the impact of any amplification attack which tries to utilize a TurboTLS server as a reflector: an attacker would have to send one UDP packet for every reflected packet generated by the server, meaning that initial requests and responses are of comparable sizes, making the amplification factor so low that it would be an ineffective use of resources. Furthermore, the UDP requests ultimately must contain a fully formed \CHELO before the server responds, limiting the amplification factor.

\subsection{TurboTLS via transparent proxying}\label{sec:discussion:proxy}

\begin{figure}[!htb]
    \centering
    \includegraphics[width=0.7\textwidth]{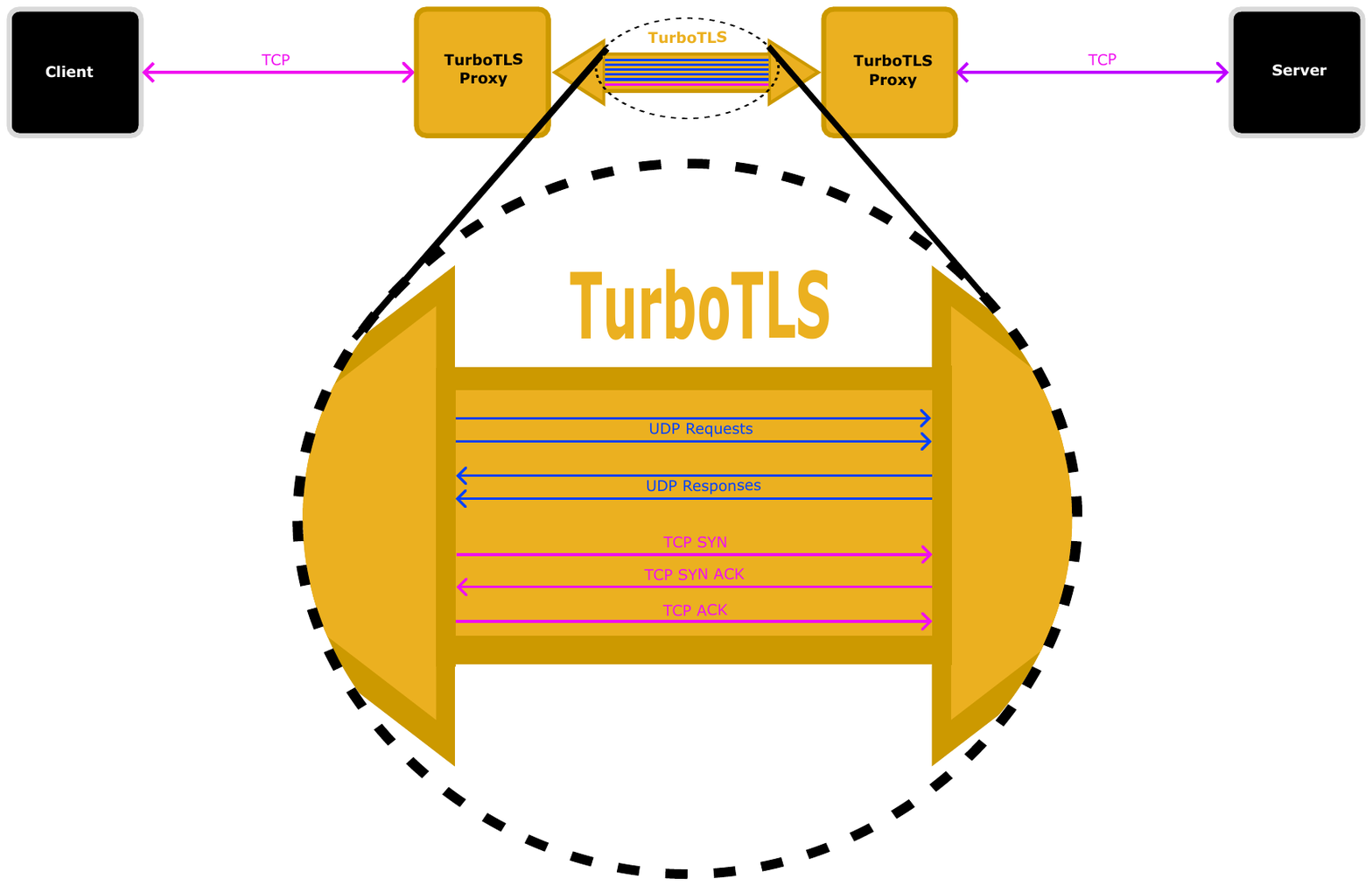}
    \caption{A client and a server each using a proxy to use TurboTLS}
    \label{fig:turbotlsproxy}
\end{figure}

Since TurboTLS does not change the contents or computations of the TLS protocol, and only changes how packets are transmitted over the network, another benefit of TurboTLS is that it can easily be implemented as a transparent proxy.  
As shown in \Cref{fig:turbotlsproxy}, a client or server (or both) could utilize a proxy to implement TurboTLS on their behalf, even for applications that only speak TLS.
A proxy could be a network device or middlebox, or even a daemon running on the same machine as the software wishing to take advantage of TurboTLS.
This greatly reduces the burden of deployment as operators can simply run a proxy rather than upgrading their networking stack.

A TurboTLS proxy would intercept TCP-based TLS communication and perform the client-based UDP fragmentation portion of TurboTLS while also setting up its own TCP connection.  
Should the UDP-based TurboTLS handshake succeed in time, the proxy would relay the \SHELO message received over the TurboTLS UDP socket to the client via the established (short-range) client-to-proxy TCP connection.
Since the proxy will be physically close to the client, the latency of establishing the TCP connection between the client and proxy will be negligible, so the client still receives the benefit from TurboTLS of not having had to wait for the TCP handshake across the long distance connection, while keeping the client completely oblivious to the fact of TurboTLS being used.
Proxies being used for performance improvements are not a new concept, and have been used for satellite communication for years \cite{rfc3135}.
Using proxies with TLS also is not new, as several software packages offer the ability to upgrade TLS versions \cite{mitmproxy,fiddler-everywhere,squid-cache,rebex}.

\subsection{TurboTLS improvements}\label{sec:discussion:variants}

We briefly mention a few alternative TurboTLS designs that may improve compatibility or can further reduce the latency assuming pre-shared state, and which may be interesting as future work.

\myparagraph{TurboTLS optimization: TurboTLS for TLS in pre-shared key mode}
Pre-shared key (PSK) mode of TLS 1.3 allows a client and server with a pre-shared symmetric key to eliminate parts of the handshake, and allows the client to optionally start sending encrypted application data its first \CtoS TLS flow, albeit without forward secrecy.
The TurboTLS technique could be applied to TLS 1.3 PSK mode, running the first TLS 1.3 PSK \CtoS and \StoC flows (including any early application data) over UDP and then switching over to TCP for the rest of the connection.  This would allow for transmission of application data on the very first \CtoS TurboTLS flow, but comes at the cost of sacrificing forward secrecy, since PSK mode does not offer it.
Early application data in both the first \CtoS and first \StoC flows would be over UDP, with only the lightweight reliability features offered by TurboTLS compared to the more extensive reliability features offered by TCP.

\myparagraph{TurboTLS optimization: TurboTLS + TLS encrypted client hello}
Encrypted client hello (ECH) \cite{ietf-tls-esni-15} is a mechanism to encrypt parts of the TLS 1.3 handshake under a semi-static server public key.
This mechanism even allows for the transmission of application data one round trip earlier, but only by sacrificing forward secrecy.
The TurboTLS approach combined with ECH could allow for transmission of application data on the very first \CtoS TurboTLS flow, at the cost of sacrificing forward secrecy.
Again, early application data flows would be over UDP with TurboTLS's lightweight reliability features compared to TCP's more extensive reliability.

\myparagraph{TurboTLS variant: UDP first stage + TLS 1.3 PSK handshake}
When the UDP and TCP payloads of TurboTLS are combined, they contain an unaltered TLS 1.3 handshake.  
However, if the TCP portion is inspected on its own, it will appear to be only a part of a TLS handshake, and there is the potential that this could cause compatibility problems for some middleboxes/firewalls/interceptors.
An alternative would be for the TLS handshake to terminate after the UDP portion of TurboTLS is completed, use the TLS keying material exporter paradigm to output a shared secret between the client and the server, and then use that shared secret as a pre-shared key in a TLS 1.3 PSK mode handshake over the TCP connection.  
This still maintains the RTT and latency improvements offered by TurboTLS, but ensures that the data within the TCP payloads are a fully standards-compliant TLS 1.3 PSK handshake transcript, which should further reduce the risk of incompatibilities from poorly configured middleboxes.
(Note this differs from the ``TurboTLS optimization: TurboTLS for TLS in pre-shared key mode'' mentioned above: the earlier paragraph on optimization for TLS in PSK mode is about using the TurboTLS technique to split a non-forward secure PSK handshakes across UDP and TCP, whereas this paragraph's TurboTLS variant does a forward-secure handshake in the UDP first stage and then uses the output of that as a PSK in a TLS 1.3 PSK handshake.)
\section{Experimental analysis}\label{sec:experiment}

We implemented TurboTLS to compare its performance with vanilla TLS 1.3 and QUIC.  
Our preliminary proof-of-concept implements most of TurboTLS as described in \Cref{sec:design}, but not yet completely; see the ``Limitations'' paragraph below.

\myparagraph{Libraries and cryptographic algorithms}
Our implementation of TurboTLS is based on OpenSSL \cite{openssl}, using Open Quantum Safe fork of OpenSSL to provide support for post-quantum algorithms \cite{SAC:SteMos16}.
We take advantage of OpenSSL's BIO interface to have a fine control over the I/O operations, allowing us to transmit some messages over UDP instead of TCP.

In our experiments, we considered two cryptographic suites, where we varied the public key algorithms used:
\begin{itemize}
\item Elliptic curves: ECDSA signatures and ECDH ephemeral key exchange using the nistp256/secp256r1 curve.
\item Post-quantum: Dilithium2 signatures \cite{NISTPQC-R3:CRYSTALS-DILITHIUM20}, and Kyber-512 key exchange \cite{NISTPQC-R3:CRYSTALS-KYBER20}. This suite results in both the \CtoS and \StoC TLS handshake flows being fragmented.
\end{itemize}
In both cases, we used the same symmetric algorithms (AES-128 in Galois counter mode, SHA-256).
We use a single self-signed certificate, in other words, a certificate chain of length 1.

\myparagraph{Network}
We used four network configurations:
\begin{itemize}
\item Local: The client and server are in the same data center of a cloud provider, with a ping time of 480--486\,microseconds.\footnote{Our TLS/TurboTLS experimental results and QUIC experiment results were collected in different data collection sessions, so there is a small difference between ping times across the two data collection sessions, which we believe is due to natural network variation.}
\item Continental: The client was in a data centre in Paris, and the server was in a data centre in Belgium, within the same cloud provider network, joined by a network connection with an observed ping time of 4.9--5.2\,milliseconds.
\item Intercontinental1: The client was in a data centre in Paris, and the server was in a data centre in Oregon, within the same cloud provider network, joined by a network connection with an observed ping time of 132--133\,milliseconds.
\item Intercontinental2: The client was in a data centre in Paris, and the server was in a data centre in Australia, within the same cloud provider network, joined by a network connection with an observed ping time of 268--270\,milliseconds.
\end{itemize}
The only source of latency we introduce is distance, but there are many other reasons for latency to be in the hundreds of milliseconds: the number of network hops (when changing between providers, or going to end users), the server load (which can provoke waiting queues over each round-trip), and the technology of the intermediate networks (IoT, 3G, etc.). In practice, global studies show\footnote{CAIDA's Macroscopic Internet Topology Monitor \url{https://www.caida.org/catalog/software/walrus/rtt/}} that, ignoring server load or end-user delays, median connections lead to RTTs mostly between 50\,ms (US west coast to US west coast) and 200\,ms (west coast to Europe), and 90th percentile connections lead to RTTs going over 500\,ms or even 1000\,ms~\cite{huffaker2002topology}. 

We addressed servers with IP addresses so we did not incur any time for DNS resolution, and the client assumed the server supported TurboTLS without making any DNS HTTPS resource record query.

\myparagraph{Machines}
In all cases, the machines used were Linux x86\_64 cloud servers with 4 cores (8 vcores taking into account HyperThreading) of an Intel Xeon E5-2696V4 Processor with 16\,GB of RAM.

\begin{figure*}[t]
\centering
\begin{subcaptionblock}[T]{0.48\textwidth}
	\centering
	\includegraphics[width=\textwidth]{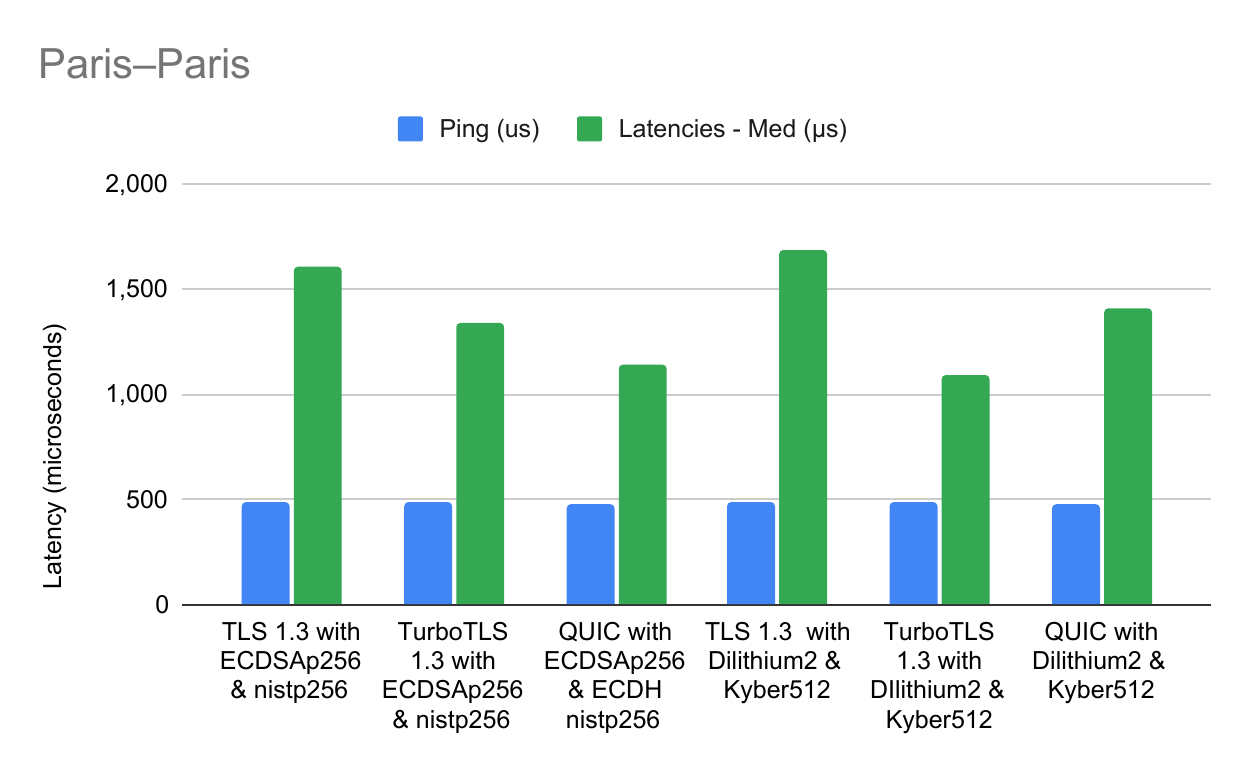}
	\caption{Local: Client and server in same datacenter}
	\label{fig:results:local}
\end{subcaptionblock}%
\begin{subcaptionblock}[T]{0.48\textwidth}
	\centering
	\includegraphics[width=\textwidth]{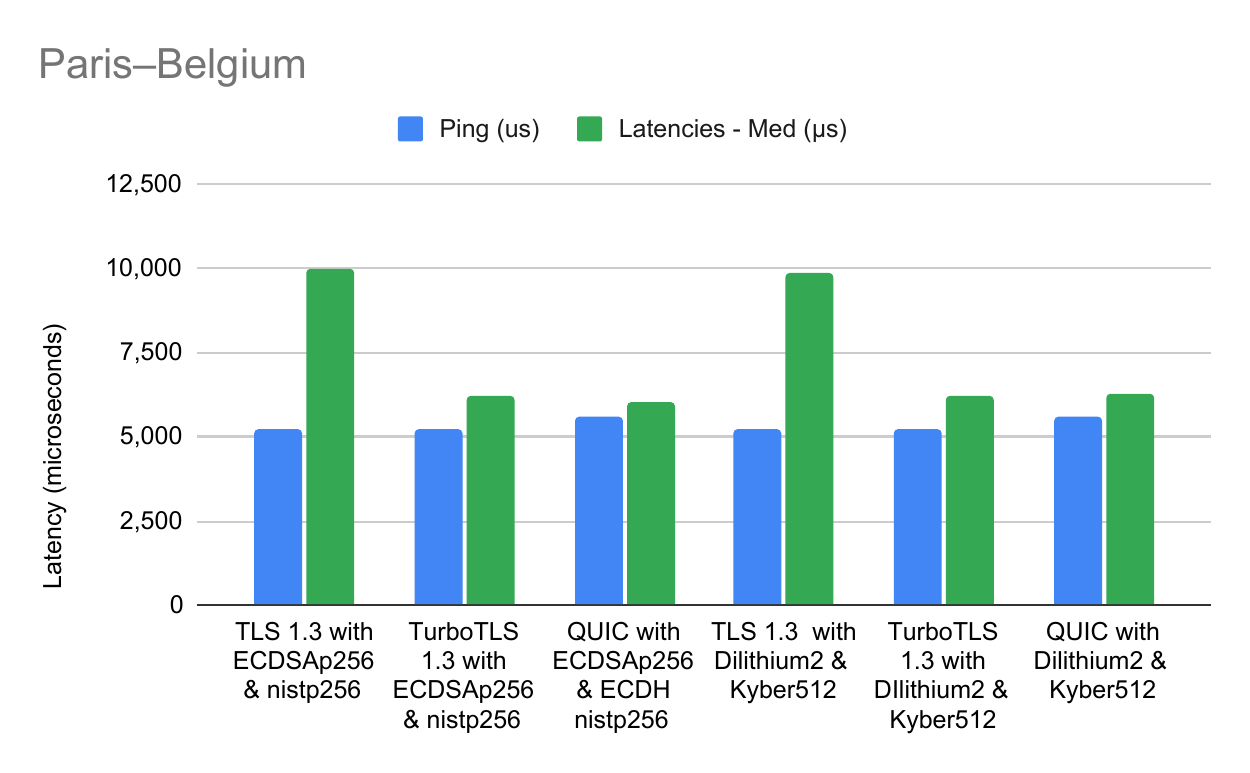}
	\caption{Continental: Client in Paris, server in Belgium}
	\label{fig:results:europe}
\end{subcaptionblock}
\begin{subcaptionblock}[T]{0.48\textwidth}
	\centering
	\includegraphics[width=\textwidth]{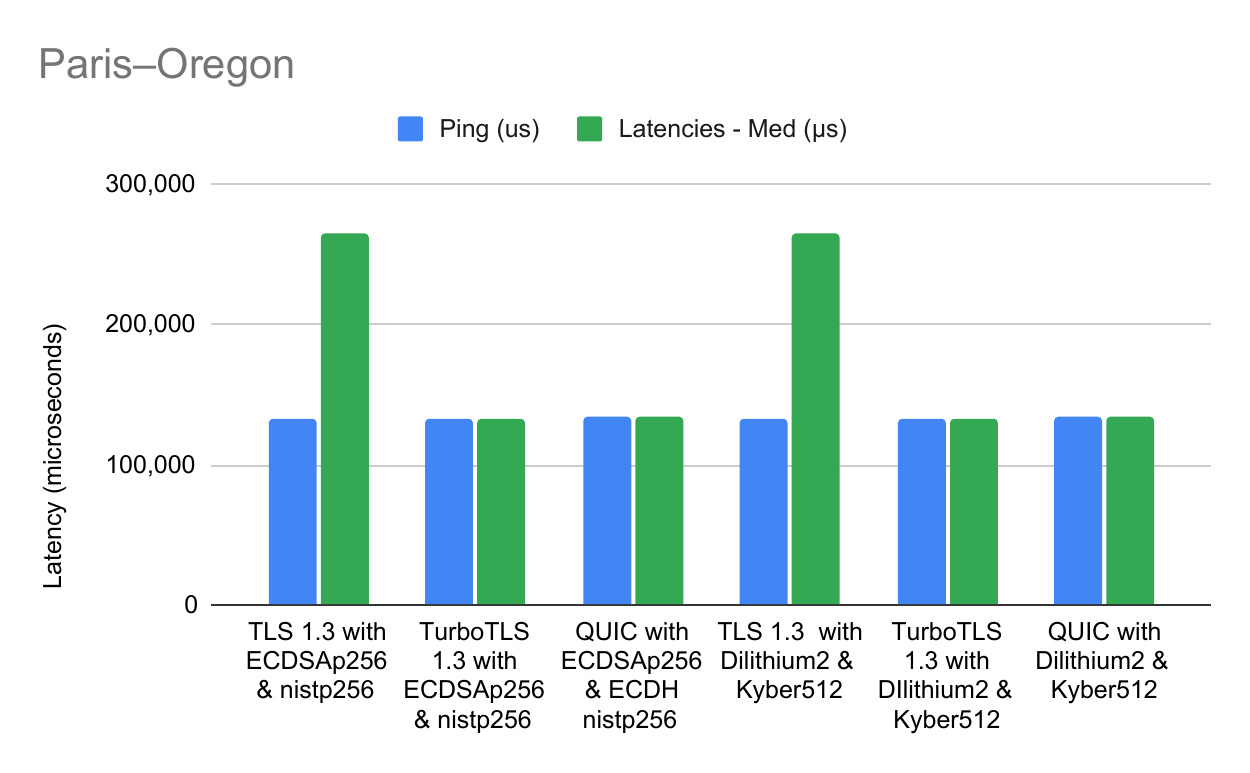}
	\caption{Intercontinental: Client in Paris, server in Oregon}
	\label{fig:results:oregon}
\end{subcaptionblock}
\begin{subcaptionblock}[T]{0.48\textwidth}
	\centering
	\includegraphics[width=\textwidth]{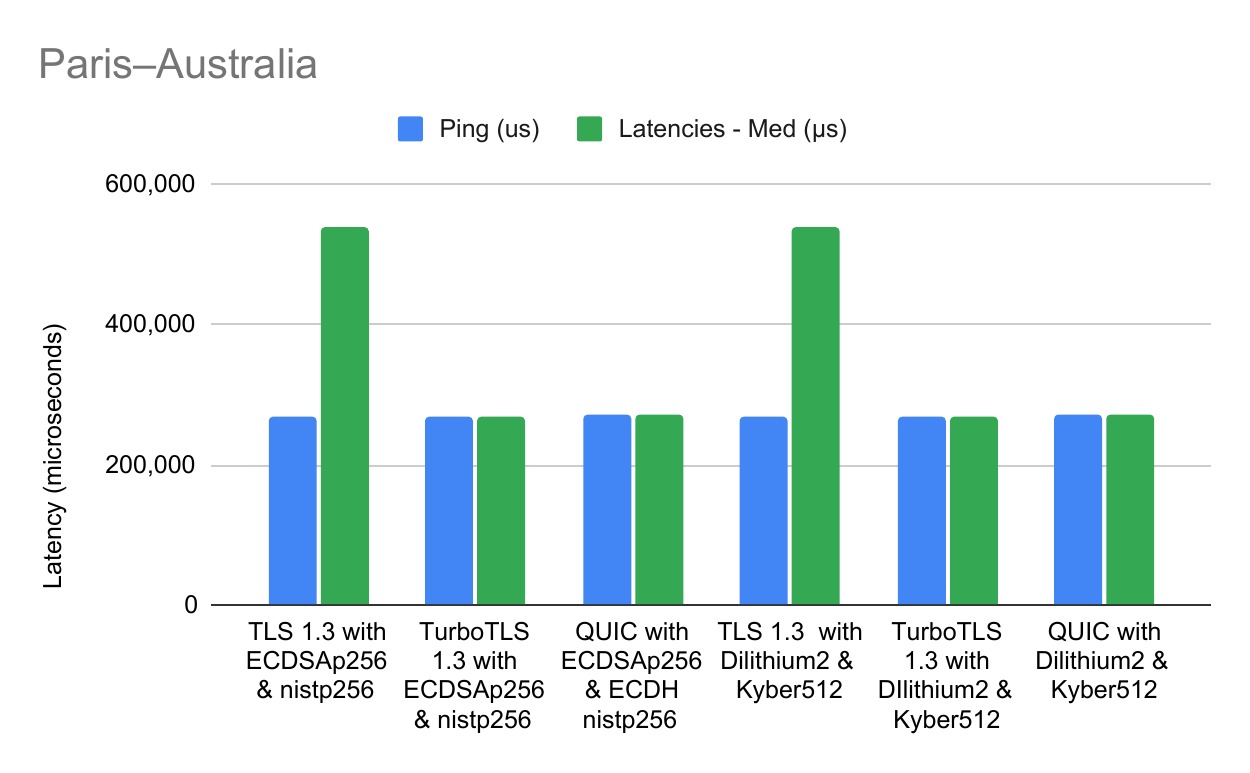}
	\caption{Intercontinental: Client in Paris, server in Australia}
	\label{fig:results:australia}
\end{subcaptionblock}

\caption{Comparing performance of TLS 1.3, TurboTLS, and QUIC in 4 network settings with 2 cryptographic suites (elliptic curves or post-quantum). Latencies reported are time in microseconds from start of connection establishment until client sends its first byte of application data.}
\label{fig:results}
\end{figure*}

\myparagraph{Results}
\Cref{fig:results} shows the results of the experiment across the two cryptographic suites and four network configurations, comparing the latencies of TLS 1.3, TurboTLS, and QUIC.  
The results reported show latencies at the 50th percentile (median) for a 10 second experiment. \Cref{tab:results} in the appendix includes the same data as well as latencies at the 90th and 99th percentile.

As expected, in long distance connections (see \Cref{fig:results:oregon}, \Cref{fig:results:australia}), where latency is primarily due to the time for information to travel between endpoints, saving one round-trip using TurboTLS approximately halves the latency compared to TLS 1.3, and TurboTLS latency behaves similarly to QUIC at both the median and 90th and 99th percentiles.

In low-latency connections (see \Cref{fig:results:local}, \Cref{fig:results:europe}), the main outcome remains that TurboTLS outperforms TLS 1.3, but there is more subtlety.
First, median latency of the 1 RTT protocols (TurboTLS and QUIC) is less than that of the 2-RTT TLS 1.3, but not fully reduced by 50\%, which is likely due to the fixed cost of the computations.
For example, in the Paris--Belgium setting (RTT of 5.2--5.6\,ms) TurboTLS and QUIC have median latency about 60\% that of TLS 1.3.
Second, latency of TLS 1.3 and TurboTLS at the 90th and 99th percentiles scales differently than QUIC.  
We believe that this is not a fundamental characteristic of TLS or TurboTLS and instead is related to the cloud environment we used (Google Cloud Platform), specifically differing behaviour of TCP versus UDP in low-latency high-throughput scenarios in hypervisors.

\myparagraph{Limitations}
Our implementation does not include, as of yet, the request-based fragmentation feature (in other words, the client only sends fragments for its actual packets, and the server responds with with as many UDP packets as needed). It also does not, as yet, fall back to TCP if the UDP packets are lost or delayed too much but, in our handful of trials, we never observed a failure of TurboTLS because of such issues (or any other reason). 
We do not expect these to substantially change performance results, as they are primarily related to compatibility.

\myparagraph{Future work} To properly evaluate the effect of request-based fragmentation, it is necessary to test over more complex heterogenous networks (those where pathways are likely to include poorly configured middleboxes and bad network connections) to fully ascertain the impact of the TLS fallback both compared to TurboTLS and to QUIC.

\section{TLS depth of websites}

The results of Section \ref{sec:experiment} are amplified  when sequential TLS connections are required: the savings from TurboTLS  gained during many connections in series could easily become human-noticeable, so we outline how this might happen.

Websites typically load resources from multiple domains (that eventually resolve to different hosts). More than one TLS connection is thus generally established by a browser to load and render a website. While most of these connections can be done in parallel, some of them can initiate only once some resources have been loaded, e.g. once some CSS of Javascript assets have been processed. As this can happen recursively, we call the {\it TLS depth} the depth of that recursion.

\begin{figure*}[htb!]
\centering
\begin{subcaptionblock}[T]{0.43\textwidth}
	\centering
	\includegraphics[width=\textwidth]{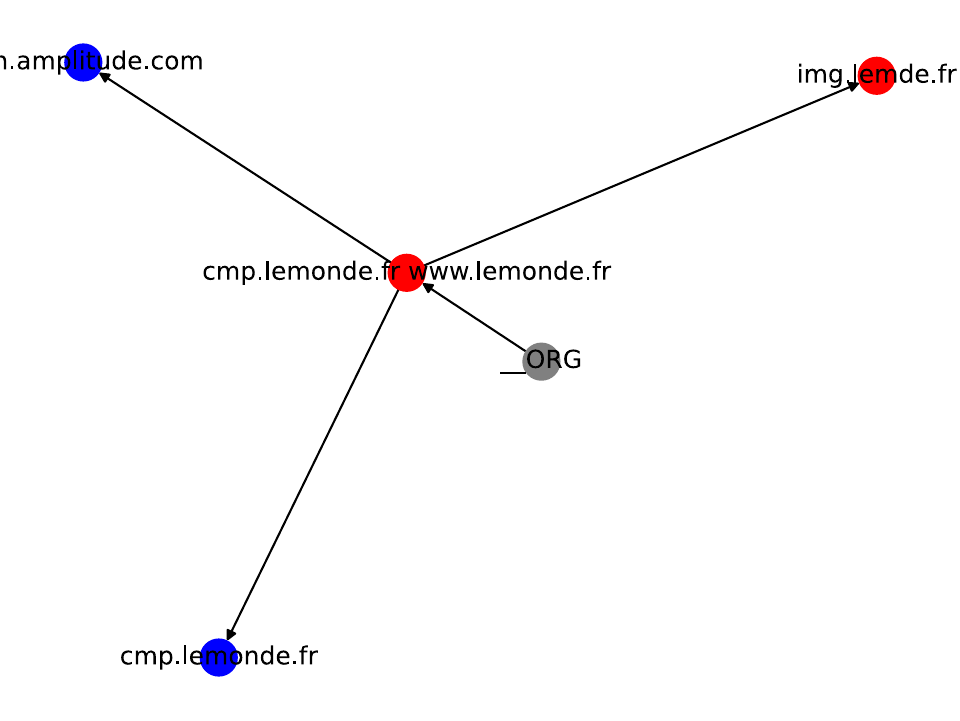}
	\caption{TLS depth of 2 for \url{ https://lemonde.fr}}
\end{subcaptionblock}
\begin{subcaptionblock}[T]{0.43\textwidth}
	\centering
	\includegraphics[width=\textwidth]{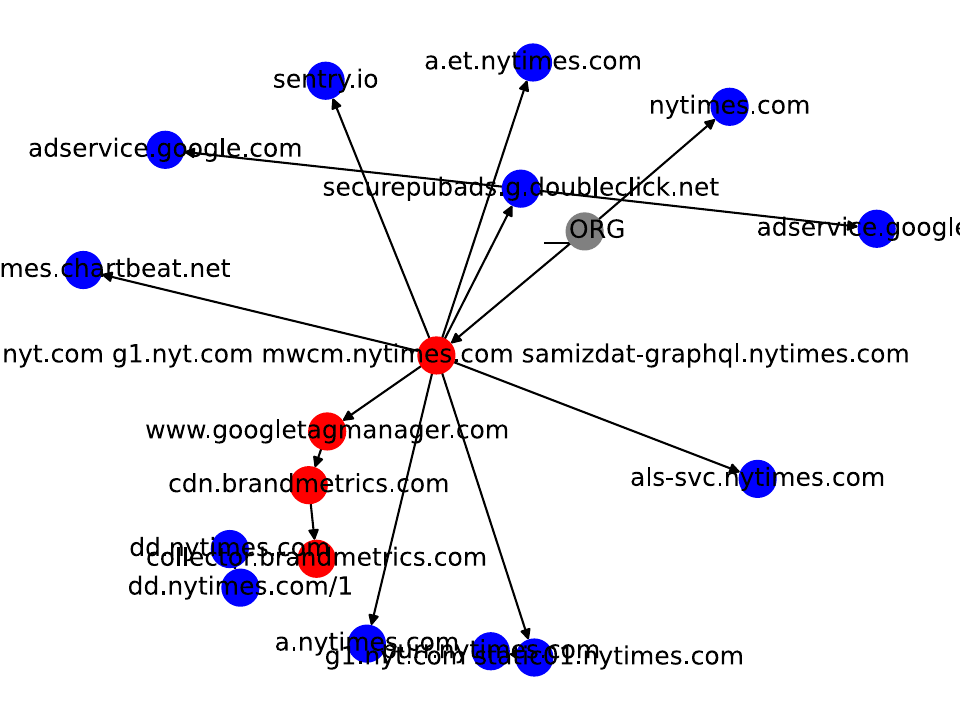}
	\caption{TLS depth of 4 for \url{ https://nytimes.com}}
\end{subcaptionblock}

\caption{Examples of TLS initiation graph for two websites, as captured on June 1, 2023 from Paris, France. Red nodes indicate one example of the longest paths (there may be more than one) leading to the TLS depth.}
\label{fig:tlsdepthexamples}
\end{figure*}

\Cref{fig:tlsdepthexamples} illustrates the TLS depth for two popular webpages as a directed graph of TLS initiations. Nodes are the hosts to which the browser made TLS connections, and an edge from host $A$ to host $B$ means that a resource downloaded from host $A$ spawned a TLS connection to host $B$. The longest path in the graph is thus the TLS depth.

\section{Identifying Usecases}\label{sec:usecases}

In this section, based upon TurboTLS's strengths and weaknesses with respect to other TLS optimizations and protocols, we provide guidance to identify applications where TurboTLS might be most profitably deployed. 

\subsection{TurboTLS as an alternative to QUIC}

QUIC is gaining significant traction and is potentially the future of internet content delivery.
QUIC solves several problems that TLS over TCP faces, including head of line blocking, mobile clients changing networks mid-connection, connection multiplexing, and improved performance for the secure connection establishment handshake. For HTTP, QUIC is undoubtedly superior to TurboTLS, if it can be deployed. 
Of course, this comes at the cost of a substantially different protocol and software stack.
Even with all of QUIC's strengths, it will be some time before QUIC is fully adopted for web traffic, and even longer before QUIC replaces TLS for the many other applications that are layered over TLS.
As TurboTLS is an optimization to TLS rather than an entirely new delivery protocol like QUIC, it may be simpler to transition to TurboTLS in order to still receive TurboTLS's handshake latency improvements.

\myparagraph{Cybersecurity concerns} Many standard network firewalls depend on information from legacy TCP sessions to detect threats. QUIC's encryption of transport information results in standard network sensors being unable to even detect QUIC. Combined with Google's frequent maintenance of QUIC, this can make security provision difficult for servers using QUIC, hence the recommendation by some providers \cite{Fortinet,Pan} to revert to (or remain on) HTTPS over TCP.

\myparagraph{Legacy applications} The issue of legacy software persists due to gnarly real-world problems of interacting applications with specific version integration requirements, subject to third party maintenance and insurance. In cases where networking upgrades such as implementing QUIC are not feasible, yet the applications running over such software are business-critical and latency sensitive, TurboTLS's proxying capabilities can be appealing due to its non-invasive implementation, and can be implemented adjacent to an application, at network gateway, or anywhere in between.

\myparagraph{Protocol simplicity} As previously mentioned, QUIC offers many great features. However, these features add additional complexity to networking stacks providing more opportunities for bugs to appear both in QUIC implementations as well as applications which use them. TurboTLS has a relatively simple design, thus making implementations less error-prone. TurboTLS can be used for those that desire additional scrutiny of QUIC implementations and their APIs.

\myparagraph{A simple alternative: Proxying TLS}
QUIC uses the TLS handshake for connection establishment, but it is integrated into the QUIC protocol and inserted into QUIC messages.
Additionally, QUIC performs some cryptographic computations differently:
in particular, as part of key derivation, QUIC provides different label strings to the HKDF compared to TLS, so the session key derived in QUIC will differ from that of TLS \cite{rfc9001}.
Consequently, without being given additional keying material, it is not possible to implement a proxy that would transparently convert a TCP-based TLS connection to a QUIC connection.
As noted in Section~\ref{sec:discussion:proxy}, a TurboTLS proxy, can merely pass TLS handshake and record layer messages back and forth across the appropriate UDP and TCP sockets without needing to know any additional keying material. Thus, whatever the motivation (security concerns, the complexity of migrating a given legacy application, or protocol complexity), users may want to replace TLS with TurboTLS, or just use a TurboTLS proxy instead of migrating to QUIC.

\subsection{Generic Turbo Transport Situations}

In some situations, all traffic transported by TCP could benefit from the underlying ideas of TurboTLS so that for any TCP connection, one round trip is removed (by a transparent proxy). We coin such an approach TurboTCP.

\myparagraph{Traffic going through an internet provider core-network}
When traffic from a client regularly follows the same path: enter a core-network, use the network to get closer to the destination, and exit such a network, it would be possible to use TurboTCP proxies that significantly improve the user experience. 

\myparagraph{High-latency satellite communications}
Proxying TCP into TurboTCP can also be systematically done by a device/OS. For example this can be beneficial with a satellite phone, that would avoid the very significant RTTs observed in such communications.

\myparagraph{Internet of Things}
In the Internet of Things, energy is of the essence, having lower latencies can also be seen as an approach to limit the time in which communicating devices are online and using energy.\

In all these settings, QUIC is clearly not an option as we are accelerating a wide span of protocols that are transported via TCP. The simplicity of proxying transparently TCP can be very attractive.

\ifanonymousversion
\else
\section{Acknowledgements}

D.S. and J.G. were supported in part by Natural Sciences and Engineering Research Council of Canada (NSERC) Discovery grant RGPIN-2022-03187.
\fi

\bibliographystyle{splncs04}

\begin{thebibliography}{10}
\providecommand{\url}[1]{\texttt{#1}}
\providecommand{\urlprefix}{URL }
\providecommand{\doi}[1]{https://doi.org/#1}

\bibitem{chrome-har-capturer}
et~al., A.C.: {Chrome HAR capturer},
  \url{https://github.com/cyrus-and/chrome-har-capturer}

\bibitem{syncookies}
Bernstein, D.J.: {SYN cookies}, \url{http://cr.yp.to/syncookies.html}

\bibitem{CAIDA}
CAIDA: {Round-Trip Time Internet Measurements from CAIDA's Macroscopic Internet
  Topology Monitor}. \url{https://www.caida.org/catalog/software/walrus/rtt/},
  accessed: 2023-06-29

\bibitem{MiddleboxesVsQuic}
Chaudhary, S., Sachdeva, P., Mondal, A., Chakraborty, S., Maity, M.: {YouTube}
  over {G}oogle's {QUIC} vs internet middleboxes: A tug of war between protocol
  sustainability and application {QoE} (2022),
  \url{https://arxiv.org/abs/2203.11977}

\bibitem{rfc7413}
Cheng, Y., Chu, J., Radhakrishnan, S., Jain, A.: {TCP Fast Open}. RFC 7413 (Dec
  2014). \doi{10.17487/RFC7413}

\bibitem{mitmproxy}
Cortesi, A., Hils, M., Kriechbaumer, T.: {mitmproxy},
  \url{https://mitmproxy.org/}

\bibitem{Fortinet}
Fortinet: {FortiGate firewall blog}.
  \url{https://community.fortinet.com/t5/FortiGate/Technical-Tip-Block-QUIC-Protocol/ta-p/197661#:~:text=Go\%20to\%20the\%20Application\%20Control,port\%2080\%20and\%20port\%20443.},
  accessed: 2023-06-29

\bibitem{arxiv.2211.14196}
Goertzen, J., Stebila, D.: Post-quantum signatures in {DNSSEC} via
  request-based fragmentation. arXiv (Nov 2022).
  \doi{10.48550/ARXIV.2211.14196}

\bibitem{rfc3135}
Griner, J., Border, J., Kojo, M., Shelby, Z.D., Montenegro, G.: {Performance
  Enhancing Proxies Intended to Mitigate Link-Related Degradations}. RFC 3135
  (Jun 2001). \doi{10.17487/RFC3135},
  \url{https://www.rfc-editor.org/info/rfc3135}

\bibitem{huffaker2002topology}
Huffaker, B., Plummer, D., Moore, D., Claffy, K.: Topology discovery by active
  probing. In: Symposium on Applications and the Internet (SAINT) 2002
  Workshops. pp. 90--96. IEEE (2002)

\bibitem{rfc9000}
Iyengar, J., Thomson, M.: {QUIC: A UDP-Based Multiplexed and Secure Transport}.
  RFC 9000 (May 2021). \doi{10.17487/RFC9000}

\bibitem{EUROSP:KraWee16}
Krawczyk, H., Wee, H.: The {OPTLS} protocol and {TLS} 1.3. In: IEEE European
  Symposium on Security and Privacy (EuroS\&P) 2016. pp. 81--96. IEEE (2016).
  \doi{10.1109/EuroSP.2016.18}

\bibitem{rfc9312}
Kühlewind, M., Trammell, B.: {Manageability of the QUIC Transport Protocol}.
  RFC 9312 (Sep 2022). \doi{10.17487/RFC9312},
  \url{https://www.rfc-editor.org/info/rfc9312}

\bibitem{agl-tls-snapstart-00}
Langley, A.: {Transport Layer Security (TLS) Snap Start}. Internet-Draft
  draft-agl-tls-snapstart-00, Internet Engineering Task Force (Jun 2010),
  \url{https://datatracker.ietf.org/doc/draft-agl-tls-snapstart/00/}

\bibitem{NISTPQC-R3:CRYSTALS-DILITHIUM20}
Lyubashevsky, V., Ducas, L., Kiltz, E., Lepoint, T., Schwabe, P., Seiler, G.,
  Stehl{\'e}, D., Bai, S.: {CRYSTALS-DILITHIUM}. Tech. rep., {N}ational
  {I}nstitute of {S}tandards and {T}echnology (2020), available at
  \url{https://csrc.nist.gov/projects/post-quantum-cryptography/round-3-submissions}

\bibitem{Pan}
Networks, P.A.: {Palo Alto Networks customer support portal}.
  \url{https://knowledgebase.paloaltonetworks.com/KCSArticleDetail?id=kA10g000000ClarCAC},
  accessed: 2023-06-29

\bibitem{CCS:PZSBL13}
Petullo, W.M., Zhang, X., Solworth, J.A., Bernstein, D.J., Lange, T.:
  {MinimaLT}: minimal-latency networking through better security. In: Sadeghi,
  A.R., Gligor, V.D., Yung, M. (eds.) ACM CCS 2013. pp. 425--438. {ACM} Press
  (Nov 2013). \doi{10.1145/2508859.2516737}

\bibitem{fiddler-everywhere}
{Progress Software Corporation}: {Fiddler Everywhere},
  \url{https://docs.telerik.com/fiddler-everywhere/security}

\bibitem{squid-cache}
project, T.S.: {Squid Web Cache}, \url{http://www.squid-cache.org/}

\bibitem{rebex}
{REBEX ČR s.r.o.}: {Rebex TLS Proxy}, \url{https://www.rebex.net/tls-proxy/}

\bibitem{rfc8446}
Rescorla, E.: {The Transport Layer Security (TLS) Protocol Version 1.3}. RFC
  8446 (Aug 2018). \doi{10.17487/RFC8446}

\bibitem{ietf-tls-ctls-06}
Rescorla, E., Barnes, R., Tschofenig, H., Schwartz, B.M.: {Compact TLS 1.3}.
  Internet-Draft draft-ietf-tls-ctls-06, Internet Engineering Task Force (Jul
  2022), \url{https://datatracker.ietf.org/doc/draft-ietf-tls-ctls/06/}

\bibitem{rfc5246}
Rescorla, E., Dierks, T.: {The Transport Layer Security (TLS) Protocol Version
  1.2}. RFC 5246 (Aug 2008). \doi{10.17487/RFC5246}

\bibitem{ietf-tls-esni-15}
Rescorla, E., Oku, K., Sullivan, N., Wood, C.A.: {TLS Encrypted Client Hello}.
  Internet-Draft draft-ietf-tls-esni-15, Internet Engineering Task Force (Oct
  2022), \url{https://datatracker.ietf.org/doc/draft-ietf-tls-esni/15/}

\bibitem{rfc9147}
Rescorla, E., Tschofenig, H., Modadugu, N.: {The Datagram Transport Layer
  Security (DTLS) Protocol Version 1.3}. RFC 9147 (Apr 2022).
  \doi{10.17487/RFC9147}

\bibitem{rfc7924}
Santesson, S., Tschofenig, H.: {Transport Layer Security (TLS) Cached
  Information Extension}. RFC 7924 (Jul 2016). \doi{10.17487/RFC7924}

\bibitem{NISTPQC-R3:CRYSTALS-KYBER20}
Schwabe, P., Avanzi, R., Bos, J., Ducas, L., Kiltz, E., Lepoint, T.,
  Lyubashevsky, V., Schanck, J.M., Seiler, G., Stehl{\'e}, D.:
  {CRYSTALS-KYBER}. Tech. rep., {N}ational {I}nstitute of {S}tandards and
  {T}echnology (2020), available at
  \url{https://csrc.nist.gov/projects/post-quantum-cryptography/round-3-submissions}

\bibitem{ietf-dnsop-svcb-https-11}
Schwartz, B.M., Bishop, M., Nygren, E.: {Service binding and parameter
  specification via the DNS (DNS SVCB and HTTPS RRs)}. Internet-Draft
  draft-ietf-dnsop-svcb-https-11, Internet Engineering Task Force (Oct 2022),
  \url{https://datatracker.ietf.org/doc/draft-ietf-dnsop-svcb-https/11/}

\bibitem{TISSEC:ShaBonRes04}
Shacham, H., Boneh, D., Rescorla, E.: Client-side caching for {TLS}. ACM Trans.
  Inf. Syst. Secur.  \textbf{7}(4),  553–575 (Nov 2004).
  \doi{10.1145/1042031.1042034}

\bibitem{NDSS:SikKamDev20}
Sikeridis, D., Kampanakis, P., Devetsikiotis, M.: Post-quantum authentication
  in {TLS} 1.3: {A} performance study. In: NDSS~2020. The Internet Society (Feb
  2020)

\bibitem{rfc6013}
Simpson, W.A.: {TCP Cookie Transactions (TCPCT)}. RFC 6013 (Jan 2011).
  \doi{10.17487/RFC6013}

\bibitem{song-atr-large-resp-03}
Song, L., Wang, S.: {ATR: Additional Truncation Response for Large DNS
  Response}. Internet-Draft draft-song-atr-large-resp-03, Internet Engineering
  Task Force (Mar 2019),
  \url{https://datatracker.ietf.org/doc/draft-song-atr-large-resp/03/}

\bibitem{SAC:SteMos16}
Stebila, D., Mosca, M.: Post-quantum key exchange for the internet and the open
  quantum safe project. In: Avanzi, R., Heys, H.M. (eds.) SAC 2016. {LNCS},
  vol. 10532, pp. 14--37. Springer, Heidelberg (Aug 2016).
  \doi{10.1007/978-3-319-69453-5_2}

\bibitem{openssl}
{The OpenSSL Project}: {OpenSSL} version 1.1.1s (Nov 2022),
  \url{https://www.openssl.org}

\bibitem{rfc9001}
Thomson, M., Turner, S.: {Using TLS to Secure QUIC}. RFC 9001 (May 2021).
  \doi{10.17487/RFC9001}, \url{https://www.rfc-editor.org/info/rfc9001}

\end{thebibliography}

\begin{table*}[t]
\centering
\noindent\makebox[\linewidth]{
\resizebox{\textwidth}{!}{
\begin{tabular}{llrrrrr}
\toprule
\multirow{2}{*}{\textbf{Protocol Mode}} & \multirow{2}{*}{\textbf{Signature \& Key Exchange}} & \multicolumn{1}{c}{\multirow{2}{*}{\textbf{Ping}}} & \multicolumn{1}{c}{\multirow{2}{*}{\textbf{Throughput}}} & \multicolumn{3}{c}{\textbf{Latency}}                                                                      \\
                                        &                                                     & \multicolumn{1}{c}{}                               & \multicolumn{1}{c}{}                                     & \multicolumn{1}{c}{\textbf{Median}} & \multicolumn{1}{c}{\textbf{P90}} & \multicolumn{1}{c}{\textbf{P99}} \\
                                        &                                                     & \multicolumn{1}{c}{$\mu$s}                             & \multicolumn{1}{c}{hs/sec}                               & \multicolumn{1}{c}{$\mu$s}              & \multicolumn{1}{c}{$\mu$s}           & \multicolumn{1}{c}{$\mu$s}           \\
\midrule
\multicolumn{7}{l}{\textbf{Paris–Paris}}                                                                                                                                                                                                                                                                                  \\
\midrule
TLS 1.3                                 & ECDSAp256 \& ECDH nistp256                          & 486                                                & 2,313                                                    & 1,611                               & 1,898                            & 2,315                            \\
TurboTLS 1.3                            & ECDSAp256 \& ECDH nistp256                          & 486                                                & 2,288                                                   & 1,337                               & 1,639                            & 2,396                          \\
QUIC                                 & ECDSAp256 \& ECDH nistp256                          & 480                                                &   ---\hspace{0.17cm}                                                 & 1,145                               & 1,257                            & 1,395                            \\
TLS 1.3                                 & Dilithium2 \& Kyber512                              & 486                                                & 2,263                                                    & 1,690                               & 2,080                            & 2,664                            \\
TurboTLS 1.3                            & Dilithium2 \& Kyber512                              & 486                                                & 2,487                                                    & 1,096                                 & 1,400                            & 2,513                            \\
QUIC                                 & Dilithium2 \& Kyber512                          & 480                                                &   ---\hspace{0.17cm}                                                  & 1,413                               & 1,590                            & 1,736                            \\
\midrule
\multicolumn{7}{l}{\textbf{Paris–Belgium}}                                                                                                                                                                                                                                                                                \\
\midrule
TLS 1.3                                 & ECDSAp256 \& ECDH nistp256                          & 5,210                                              & 399                                                      & 10,012                               & 10,802                           & 11,087                           \\
TurboTLS 1.3                            & ECDSAp256 \& ECDH nistp256                          & 5,210                                              & 553                                                      & 6,196                               & 10,164                            & 11,189                            \\
QUIC                                 & ECDSAp256 \& ECDH nistp256                          & 5,620                                                & ---\hspace{0.07cm}                                                     & 6,016                               & 6,401                            & 6,800                            \\
TLS 1.3                                 & Dilithium2 \& Kyber512                              & 5,210                                              & 401                                                      & 9,842                              & 10,829                           & 11,667                           \\
TurboTLS 1.3                            & Dilithium2 \& Kyber512                              & 5,210                                             & 562                                                      & 6,239                               & 10,264                            & 11,071                            \\
QUIC                                 & Dilithium2 \& Kyber512                          & 5,620                                                &   ---\hspace{0.07cm}                                                   & 6,275                               & 6,565                            & 6,727                            \\
\midrule
\multicolumn{7}{l}{\textbf{Paris–Oregon}}                                                                                                                                                                                                                                                                                 \\
\midrule
TLS 1.3                                 & ECDSAp256 \& ECDH nistp256                          & 132,021                                            & 15                                                       & 264,618                             & 265,417                          & 266,728                          \\
TurboTLS 1.3                            & ECDSAp256 \& ECDH nistp256                          & 132,021                                            & 30                                                       & 132,754                             & 133,199                          & 134,053                          \\
QUIC                                 & ECDSAp256 \& ECDH nistp256                          & 133,525                                                &  ---                                                    & 134,363                               & 134,652                            & 134,788                            \\
TLS 1.3                                 & Dilithium2 \& Kyber512                              & 132,021                                            & 15                                                       & 265,430                             & 266,268                          & 267,505                          \\
TurboTLS 1.3                            & Dilithium2 \& Kyber512                              & 132,021                                            & 29                                                       & 132,562                             & 132,986                          & 133,479                          \\
QUIC                                 & Dilithium2 \& Kyber512                          & 133,525                                                &   ---                                                  & 134,479                               & 134,775                            & 134,897                            \\
\midrule
\multicolumn{7}{l}{\textbf{Paris–Australia}}                                                                                                                                                                                                                                                                              \\
\midrule
TLS 1.3                                 & ECDSAp256 \& ECDH nistp256                          & 268,157                                            & 7                                                        & 538,426                             & 539,390                          & 539,787                          \\
TurboTLS 1.3                            & ECDSAp256 \& ECDH nistp256                          & 268,157                                            & 14                                                       & 269,753                             & 270,321                          & 273,781                          \\
QUIC                                 & ECDSAp256 \& ECDH nistp256                          & 270,453                                                & ---                                                     & 271,018                               & 271,553                           & 271,839                            \\
TLS 1.3                                 & Dilithium2 \& Kyber512                              & 268,157                                            & 7                                                        & 539,279                             & 540,387                          & 541,679                          \\
TurboTLS 1.3                            & Dilithium2 \& Kyber512                              & 268,157                                            & 14                                                       & 269,606                             & 270,026                          & 271,118                          \\
QUIC                                 & Dilithium2 \& Kyber512                          & 270,453                                                & ---                                                  & 271,262                               & 271,793                            & 272,056                            \\
\bottomrule
\end{tabular}
}
}
\caption{Performance of TLS 1.3, TurboTLS, and QUIC in 4 network settings with 2 cryptographic suites (elliptic curves or post-quantum). Latencies reported are time in microseconds from start of connection establishment until client sends its first byte of application data.}
\label{tab:results}
\end{table*}

 \begin{subappendices}
\renewcommand{\thesection}{\Alph{section}}
\section{Computing TLS depth}\label{sec:comp_tls_depth}
We consider a website is loaded once the Javascript {\tt onload} event has been fired. A website that loads all its assets from the same host has a TLS depth of one. For example, take the website {\tt https://mywebsite.com} that first loads Javascript code from {\tt https://mycdn.com}. Next, that Javascript code fetches some images from {\tt https://coolimages.com}. This illustrates how loading the webpage for the user may take many sequential TLS connections before being ready.

For computing the TLS depth of real-life websites, we need to capture all TLS requests that have been done by a browser from the moment the website starts loading up until the {\tt onload} Javascript event has been fired. Most importantly, we also need to capture the initiator of these TLS request (eg. a fetch request in Javascript code), so that we can compute the complete graph of TLS initiators.

In practice, the Google Chrome browser can be used to load a website, stop on the ${\tt onload}$ Javascript event, and dump all requests that have been done in the HTTP ARchive (HAR) format. The open-source project ${\tt chrome-har-capturer}$ \cite{chrome-har-capturer} does this automatically. The Chrome browser includes in its HAR dump an ${\tt initiator}$ field that can be used to create our TLS initiator graph. Then, using the tool, these HAR dumps can be parsed to generate TLS initiator graphs and thus compute the TLS depth of websites. \end{subappendices}
\end{document}